\documentclass{article}

\usepackage{arxiv}
\usepackage{latexsym}

\usepackage{color}
\definecolor{dkgreen}{rgb}{0,0.6,0}
\definecolor{gray}{rgb}{0.5,0.5,0.5}
\definecolor{mauve}{rgb}{0.90,0,0.51}
\definecolor{lightblue}{rgb}{0.9, 0.95, 1.0}
\definecolor{lightgray}{rgb}{0.9, 0.9, 0.9}

\usepackage[
  breaklinks=true,
  colorlinks,
  linkcolor=mauve,
  citecolor=dkgreen,
  bookmarks=false
]{hyperref}
\usepackage[utf8]{inputenc} 
\usepackage[T1]{fontenc}    
\usepackage{hyperref}       
\usepackage{url}            
\usepackage{booktabs}       
\usepackage{amsfonts}       
\usepackage{nicefrac}       
\usepackage{microtype}      
\usepackage{lipsum}
\usepackage{graphicx}
\usepackage{amsmath}
\usepackage{enumitem}
\usepackage{adjustbox}
\usepackage{multirow}
\usepackage{wrapfig}
\graphicspath{ {./images/} }

\title{AUV-Fusion: Cross-Modal Adversarial Fusion of User Interactions and Visual Perturbations Against VARS}

\author{Hai Ling${^1}$ \ \ \ \ \ Tianchi Wang${^1}$ \ \ \ \ \ Xiaohao Liu${^2}$ \ \ \ \ \ Zhulin Tao${^1}$ \ \ \ \ \ Lifang Yang${^1}$ \ \ \ \ \  Xianglin Huang${^1}$\\
${^1}$ Communication University of China  \ \ \ \ \ ${^2}$ National University of Singapore \\
\tt\small lingh@cuc.edu.cn \quad wangtianchi@cuc.edu.cn
}

\begin{document}
\maketitle
\begin{abstract}
Modern Visual-Aware Recommender Systems (VARS) exploit the integration of user interaction data and visual features to deliver personalized recommendations with high precision. However, their robustness against adversarial attacks remains largely underexplored, posing significant risks to system reliability and security. Existing attack strategies suffer from notable limitations: shilling attacks are costly and detectable, and visual-only perturbations often fail to align with user preferences. To address these challenges, we propose AUV-Fusion, a cross-modal adversarial attack framework that adopts high-order user preference modeling and cross-modal adversary generation. Specifically, we obtain robust user embeddings through multi-hop user–item interactions and transform them via an MLP into semantically aligned perturbations. These perturbations are injected onto the latent space of a pre-trained VAE within the diffusion model. By synergistically integrating genuine user interaction data with visually plausible perturbations, AUV-Fusion eliminates the need for injecting fake user profiles and effectively mitigates the challenge of insufficient user preference extraction inherent in traditional visual-only attacks. Comprehensive evaluations on diverse VARS architectures and real-world datasets demonstrate that AUV-Fusion significantly enhances the exposure of target (cold-start) items compared to conventional baseline methods. Moreover, AUV-Fusion maintains exceptional stealth under rigorous scrutiny.
\end{abstract}


\section{Introduction}

\begin{figure}[!htb]
  \centering
  \includegraphics[width=0.9\linewidth]{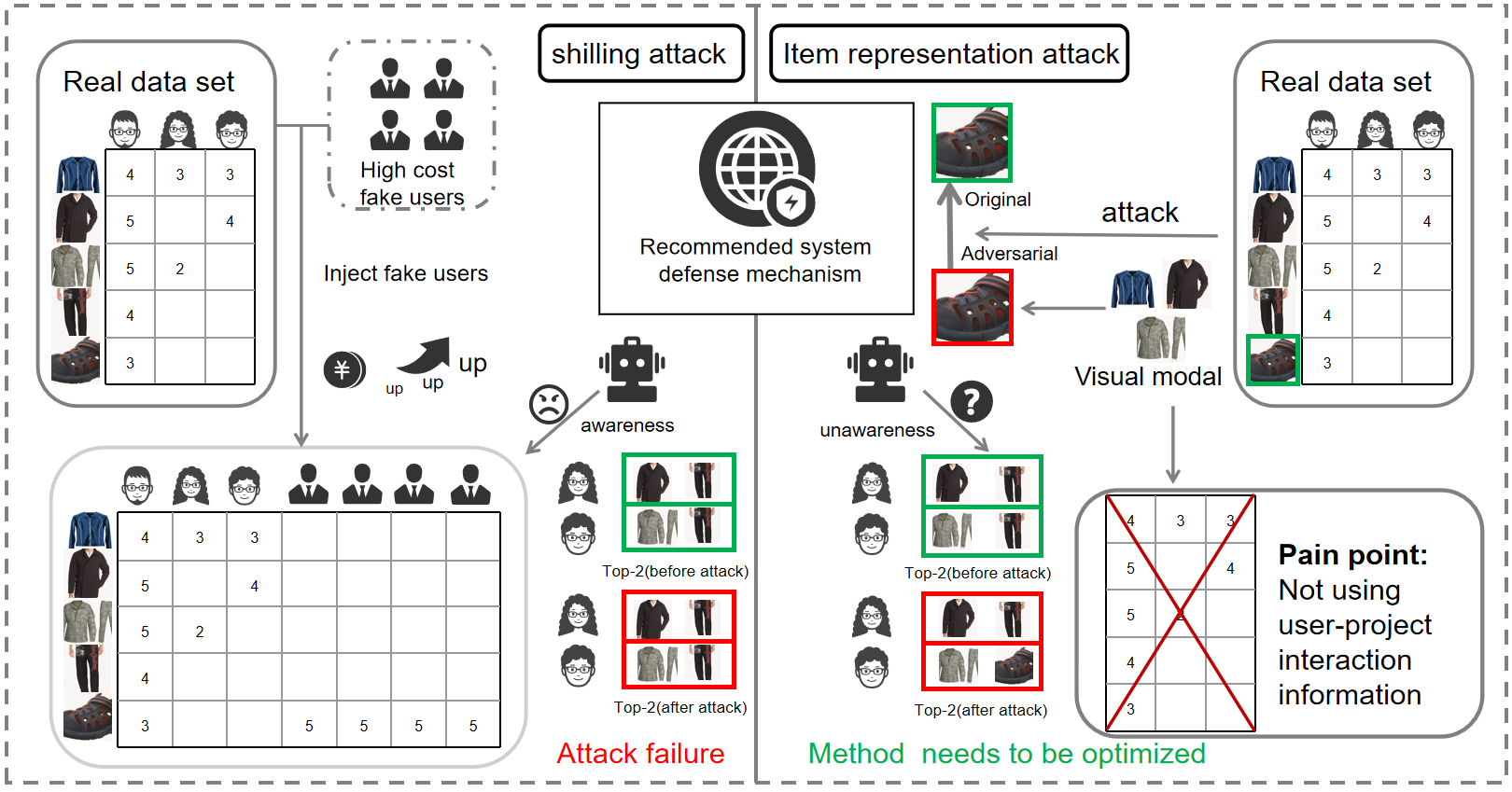}
  \caption{\textbf{Illustration of how a Visual-Aware Recommendation System faces different vulnerabilities.} Shilling attacks require numerous fake user profiles, whereas item representation attacks operate directly on item features, incurring lower costs and potentially higher stealth.}
  \label{fig:image} 
\end{figure}

With the rapid advancement of Internet technologies, recommendation systems have become a fundamental infrastructure to enhance user experience in the era of big data~\cite{covington2016deep,yuan2020parameter}. However, the exponential growth of platform users has led to increasingly sparse user-item interaction matrices, posing significant challenges to the effectiveness of recommendations~\cite{koren2009matrix}. To address this issue, modern recommendation systems try to incorporate auxiliary features for performance improvement~\cite{deldjoo2023review}. 
Particularly, Visual-Aware Recommendation System (VARS) employs item images containing rich semantic information to improve recommendation performance~\cite{he2016vbpr,yu2018aesthetic}. The integration of visual features alleviates data sparsity issues, thus demonstrating remarkable effectiveness, especially in cold-start scenarios.

Despite the benefits that visual features bring, their introduction may expose recommendation systems to potential security vulnerabilities. As shown in the Figure~\ref{fig:image}, existing studies~\cite{yang2024attacking,liu2021adversarial,chen2024adversarial} reveal that VARS is susceptible to adversarial poisoning attacks. For instance, on e-commerce platforms, merchants often seek to promote their products in recommendation rankings based on user preferences. Since platforms allow merchants to freely upload product images, some might craft adversarial images to deceive the recommendation ranking mechanisms, thus unfairly boosting product exposure and attracting more user attention. This requires in-depth research on adversarial image threats in Visual-Aware Recommendation Systems to ensure system security and fairness.

Recently, numerous studies~\cite{duan2024attacking,liu2024toda,fan2023adversarial,wu2023influence,nguyen2023poisoning,fang2018poisoning,gunes2014shilling,huang2021data} demonstrate that recommendation systems are vulnerable to \emph{shilling attacks} (\textit{a.k.a.}, data poisoning attacks), which manipulate recommendation results through fabricated user profiles. However, the effectiveness of shilling attacks is typically constrained by the high cost of generating large-scale fake user profiles and the risk of detection due to abnormal interaction patterns~\cite{10.1145/3447548.3467335,10.1145/3404835.3462914}. 
In contrast, \emph{item representation attacks} (\textit{e.g.}, adversarial image perturbations) circumvent the need for fake users by directly modifying item features, leading to lower overhead and increased stealth in real-world scenarios~\cite{liu2021adversarial}. As shown in Figure~\ref{fig:image}, item representation attacks bypass user profile creation and instead exploit visual or textual item representations, making them harder to detect with typical shilling-based defenses.

Unfortunately, existing attacks~\cite{liu2021adversarial,chen2024adversarial,yang2024attacking} on modern VARS mainly focus on constructing adversarial samples using merely single-modal information (\textit{i.e.}, transferring between images). Although these vision-oriented attacks show notable effectiveness, they fail to fully exploit the crucial user-item interaction modality information that shilling attacks typically utilize. To tackle the above problems, in this paper, we propose AUV-Fusion, a multimodal adversarial attack framework that synergistically combines user-item interaction dynamics and visual perturbations to achieve stealthy yet effective recommendation poisoning.

Our framework bridges the gap between adversarial feature manipulation and interaction-driven attacks through two synergistic modules.
\textbf{(1) High-order User Preference Modeling}: We integrate both collaborative interactions and semantic visual features through a multi-view strategy. The user–item view employs a LightGCN to propagate multi-hop interactions on the user–item graph, iteratively aggregating neighborhood information to generate robust behavioral embeddings. Concurrently, the item–item view constructs an affinity graph based on the cosine similarity of raw image features, applies KNN-based sparsity to reduce noise, and leverages graph convolution to enhance salient visual signals. By fusing the collaborative and visual representations as \( \mathbf{e}_{u} = \mathbf{e}_{u,id} + \mathbf{e}_{u,v} \), our method produces a comprehensive user embedding that effectively captures global preferences and provides a solid foundation for downstream adversarial image generation.
\textbf{(2) Cross-modal Adversary Generation}: We incorporate user-driven adversarial adjustments into the latent space of a pretrained diffusion model. The prior high-order user embeddings, which encapsulate both collaborative interactions and visual semantics, are processed through a multilayer perceptron to produce a latent-space perturbation. This perturbation is adaptively injected into the VAE-derived latent representation during the forward diffusion process and combined with controlled Gaussian noise. Deterministic DDIM sampling is then applied in the reverse process to reconstruct an adversarial image that preserves semantic consistency and visual fidelity. 
Furthermore, a hybrid loss function is employed to balance attack potency and imperceptibility; this loss comprises CLIP-based semantic preservation, SSIM for maintaining structural integrity, and a user alignment loss. In this manner, AUV-fusion effectively bridges adversarial feature manipulation with interaction-driven attacks to yield adversarial images tailored to user preferences.

Our main contributions are summarized as follows:

\begin{itemize}
    \item We propose AUV-Fusion, the first cross-modal adversarial attack framework for Visual-Aware Recommender System (VARS). It leverages user interactions to create visually plausible perturbations, addressing the need for fake user profiles, which is a longstanding drawback of traditional shilling attacks.

    \item AUV-Fusion overcomes the limited user preference extraction in visual-only attacks by integrating a high-order preference modeling module with a diffusion-based perturbation generator, producing adversarial changes that match genuine user preference for better attack effectiveness.
    
    \item Extensive evaluations on three widely used VARS architectures across real-world datasets demonstrate that AUV-Fusion significantly boosts the exposure of cold-start target items while maintaining superior stealth under advanced defense scenarios.
\end{itemize}
\section{Related Works}\label{app4}
\subsection{Visually-aware Recommender Systems}
Visual perception-aware recommendation systems have significantly enhanced recommendation performance by integrating semantically rich images into traditional collaborative filtering models. Prior to the rise of multimodal learning~\cite{radford2021learning, liu2024towards, liu2025continual}, most Visual-Aware Recommendation Systems primarily relied on image retrieval techniques. For instance, Kalantidis et al.~\cite{kalantidis2013getting} proposed a scalable method that automatically recommends products by segmenting and clustering foreground regions in single images (without metadata) to detect garment categories, followed by visual similarity-based retrieval within each category. Subsequently, Jagadeesh et al.~\cite{jagadeesh2014large} introduced a large-scale visual recommendation system employing four data-driven models (Complementary Nearest Neighbor Consensus, Gaussian Mixture Model, Texture-Agnostic Retrieval, and Markov Chain LDA) to recommend complementary items based on input images. They also released the Fashion-136k dataset, a large-scale annotated fashion image corpus.

Following the success of Convolutional Neural Networks~\cite{he2016deep,simonyan2014very} and deep learning-based collaborative filtering~\cite{rendle2012bpr,he2017neural}, researchers began exploring the integration of visual features into user-item interactions~\cite{he2016vbpr, mcauley2015image, liu2022elimrec, tao2022self}. IBR~\cite{mcauley2015image} pioneered the modeling of human-perceived visual relationships between items, such as identifying substitutes or complements, extending beyond mere visual similarity to capture human perceptions of visual compatibility. Subsequent models like VBPR~\cite{he2016vbpr} and Fashion DNA~\cite{bracher2016fashion} further incorporated visual information into collaborative filtering frameworks by jointly leveraging latent factors and visual features. Notably, VBPR~\cite{he2016vbpr} represents the first work to integrate pre-trained CNN visual features into collaborative filtering, emphasizing visual information's critical role in fashion recommendations. Differing from VBPR's pre-trained feature extraction, Kang et al.~\cite{kang2017visually} proposed DVBPR, an end-to-end framework that jointly trains CNN-based visual feature extraction and recommendation models using raw images as input. Despite their effectiveness in improving recommendation accuracy and alleviating cold-start problems, Visual-Aware Recommendation Systems introduce novel security risks. Tang et al.~\cite{tang2019adversarial} revealed vulnerabilities caused by maliciously manipulated item images and proposed AMR, a robustness-oriented visual recommendation model, to address these threats.

However, existing defenses like AMR primarily focus on visual perturbations and overlook the cross-modal nature of modern VARS, where adversarial risks may arise from the interplay of visual and interaction data. Our work addresses this gap by proposing a unified attack framework that jointly exploits both modalities, exposing vulnerabilities beyond visual-only perturbations.

\subsection{Item Representation Attacks}
Although recommendation systems have achieved remarkable success in real-world platforms, growing evidence~\cite{duan2024attacking,liu2024toda,fan2023adversarial,wu2023influence,nguyen2023poisoning,fang2018poisoning,gunes2014shilling,huang2021data} reveals that modern deep learning-based recommendation systems exhibit significant vulnerabilities to carefully crafted adversarial attacks. These attacks, commonly termed shilling attacks, manipulate training data by injecting fake users who assign artificially high/low ratings to target items, thereby distorting recommendation outcomes. Such manipulation not only degrades recommendation accuracy by promoting irrelevant or low-quality items but also undermines platform profitability and user experience. However, while effective, traditional shilling attacks face scalability limitations due to the high costs associated with injecting sufficient fake user profiles.

Building upon shilling attacks~\cite{9806457, liu2024toda}, only a few studies~\cite{liu2021adversarial,chen2024adversarial,yang2024attacking,11060893} have explored adversarial attacks against modern visual perception-aware recommendation systems to systematically assess their robustness and defense mechanisms~\cite{10562343,10423793}. Liu et al.~\cite{liu2021adversarial} pioneered this direction by proposing item representation attacks (INSA, EXPA, and c-SEMA), the first framework to deceive recommendation systems through perturbations in visual content. Chen et al.~\cite{chen2024adversarial} advanced this paradigm by clustering images using K-means, and then aligning the features of the target object with those of the most popular object in the corresponding cluster to achieve an effective attack. The most recent work by Yang et al.~\cite{yang2024attacking} introduced a style-transfer-based approach that transplants visual styles from popular items to target items, effectively boosting their recommendation exposure while enabling cross-system attack generalization. Despite these advancements, current methods remain constrained by their reliance on single-modality information fusion, and the perceptual quality of generated adversarial images still requires substantial improvement.

These prior works have demonstrated the feasibility of attacking visual recommendation systems using visual perturbations.   However, they primarily focus on manipulating visual features in isolation or through simple cross-modal alignments.   A significant limitation is their inability to effectively leverage the rich information embedded in user preferences to guide the attack, often leading to suboptimal results or easily detectable perturbations.   In contrast, our proposed method, AUV-Fusion, addresses these limitations by introducing a novel cross-modal adversarial attack framework.   AUV-Fusion leverages high-order user preference modeling to generate semantically aligned perturbations that are injected into the latent space of a VAE within a diffusion model.   This approach not only eliminates the need for costly fake user profile injection inherent in traditional shilling attacks but also overcomes the perceptual limitations of visual-only attacks by synergistically integrating genuine user interaction data with visually plausible perturbations, thereby achieving stealthier and more effective attacks
\section{Problem Formulation}
Visual-Aware Recommendation System (VARS) leverages both user-item interactions and item visual content for personalized recommendations. We formalize the system through three components.

\textbf{Interaction graph.} Let $G = (\mathcal{U},\mathcal{I},\mathcal{E})$ be a bipartite graph, where $\mathcal{U}$ and $\mathcal{I}$ denote disjoint sets of users and items, respectively; $\mathcal{E} \in |\mathcal{U}| \times |\mathcal{I}|$ represents observed interactions (\textit{e.g.}, clicks, purchases); Each edge $(u,i) \in \mathcal{E}$ is associated with a binary label $y_{u,i} \in \{0,1 \}$, where $y_{u,i}=1$ indicates interaction.

\textbf{Visual content.} Each item $i \in \mathcal{I}$ has an associated image $v_i \in \mathbb{R}^{H \times W \times C}$ (height $H$, width $W$, channels $C$). The system extracts visual features $\mathbf{f}_i = \Phi(v_i)\in \mathbb{R}^d$ using a pretrained encoder $\Phi$ (\textit{e.g.}, ResNet~\cite{he2016deep}).

\textbf{Recommendation mechanism.} User/item embeddings $\mathbf{e}_u,\mathbf{e}_i\in\mathbb{R}^d$ are learned from graph $G$ and $\{\mathbf{f}_i\}$. Recommendation scores are computed as $r_{u,i} = \text{sim}(\mathbf{e}_{u},\mathbf{e}_{i})$, where $\text{sim}$ is a similarity function (\textit{e.g.}, dot product). Subsequently, the top-$k$ recommendations for a user $u$, denoted as $\mathcal{R}_u^{(k)}=\text{top-}k\ r_{u,i}$, are generated by ranking the recommendation scores accordingly.

Let $\mathcal{I}_{\text{cold}} \subseteq \mathcal{I}$ denote the \textit{cold-start item set}, defined as the $K$ items with the fewest interactions:
$$\mathcal{I}_{\text{cold}} = \mathop{\underset{i \in \mathcal{I}}{\text{argmin}-K}}\left( \text{deg}(i) \right)$$ where $\text{deg}(i) = \sum_{u \in \mathcal{U}}y_{u,i}$ denotes the interaction degree of item $i$, and $\text{argmin-}K$ selects the $K$ items with the lowest degrees. In line with the setting in \cite{yang2024attacking}, we set $K=10$.
\subsection{Attack Objective}
The adversary aims to manipulate images $\{v_i\}$ of target items $\mathcal{I}_{\text{attack}} \subseteq \mathcal{I}_{\text{cold}}$ into adversarial versions $\{v_i^{\text{attack}}\}$, such that:
(1) \textbf{Exposure boost:} Increase the hit rate of $\mathcal{I}_{\text{attack}}$ in $\text{top-}k$ recommendations; and 
(2) \textbf{Visual plausibility: } $v_i^{\text{attack}}$ remains perceptually similar to $v_i$.
Formally, the attack solves:
\begin{equation}
    \mathop{\underset{\{v_i^{\text{attack}}\}}\max}\underbrace{\frac{1}{|\mathcal{I}_{\text{attack}}|}\sum_{i \in \mathcal{I}_{\text{attack}}} \text{HR}_i@k}_{\text{Exposure Gain}}-\lambda \underbrace{\frac{1}{|\mathcal{I}_{\text{attack}}|}\sum_{i \in \mathcal{I}_{\text{attack}}}\mathcal{D}(v_i^{\text{attack}},v_i)}_{\text{Detection Risk}}.
\end{equation}
where $\text{HR}_{i}@k=\frac{1}{|\mathcal{U}|}\sum_{u \in \mathcal{U}}\mathbb{I}(i \in \mathcal{R}_u^{(\text{attack})(k)})$. $\mathcal{D}(\cdot)$ quantifies perceptual distortion and $\lambda$ balances stealth \textit{v.s.} effectiveness.
\subsection{Attack Capability}
The adversary operates under a strict black-box threat model with the following capabilities and constraints:

\textbf{No access to system internals.}
Under this threat model, the adversary is completely model-agnostic, meaning it cannot access the recommendation model's architecture, parameters, or training procedures. Moreover, it is feature encoder$-$blind, having no knowledge of the visual feature extractor $\Phi$ that maps images $v_i$ to embeddings $\mathbf{f}_i$.

\textbf{Partial interaction data.}
The adversary is granted access to the interaction histories of a subset of users, denoted as $\mathcal{U}_{\text{obs}} \subseteq \mathcal{U}$, where the number of observed users is given by $|\mathcal{U}_{\text{obs}}| = \lfloor p|\mathcal{U}| \rfloor$ for some $p \in [0,1]$. In addition, the adversary obtains the observed interaction matrix, $\mathcal{E}_{\text{obs}}$, which is of dimension $|\mathcal{U}_{\text{obs}}| \times |\mathcal{I} \setminus \mathcal{I}_{\text{cold}}|$ and records the interactions between these observed users and the items that are not part of the cold-start set.

\textbf{Full visual access to target items.}
The adversary has read privileges to obtain the original images: 
$$
\{v_i \mid i \in I_{\text{attack}} \cup \{ i \mid (u,i) \in \mathcal{E}_{\text{obs}} \}\},
$$
which includes all target cold-start items as well as those items featured in the observed interactions. In addition, the adversary holds write privileges that allow it to deploy the perturbed images \(\{v_i^{\text{attack}}\}\) for every item \(i \in \mathcal{I}_{\text{attack}}\).
\section{AUV-Fusion}\label{4}
\begin{figure*}[t]
  \centering
  \includegraphics[width=0.94\linewidth]{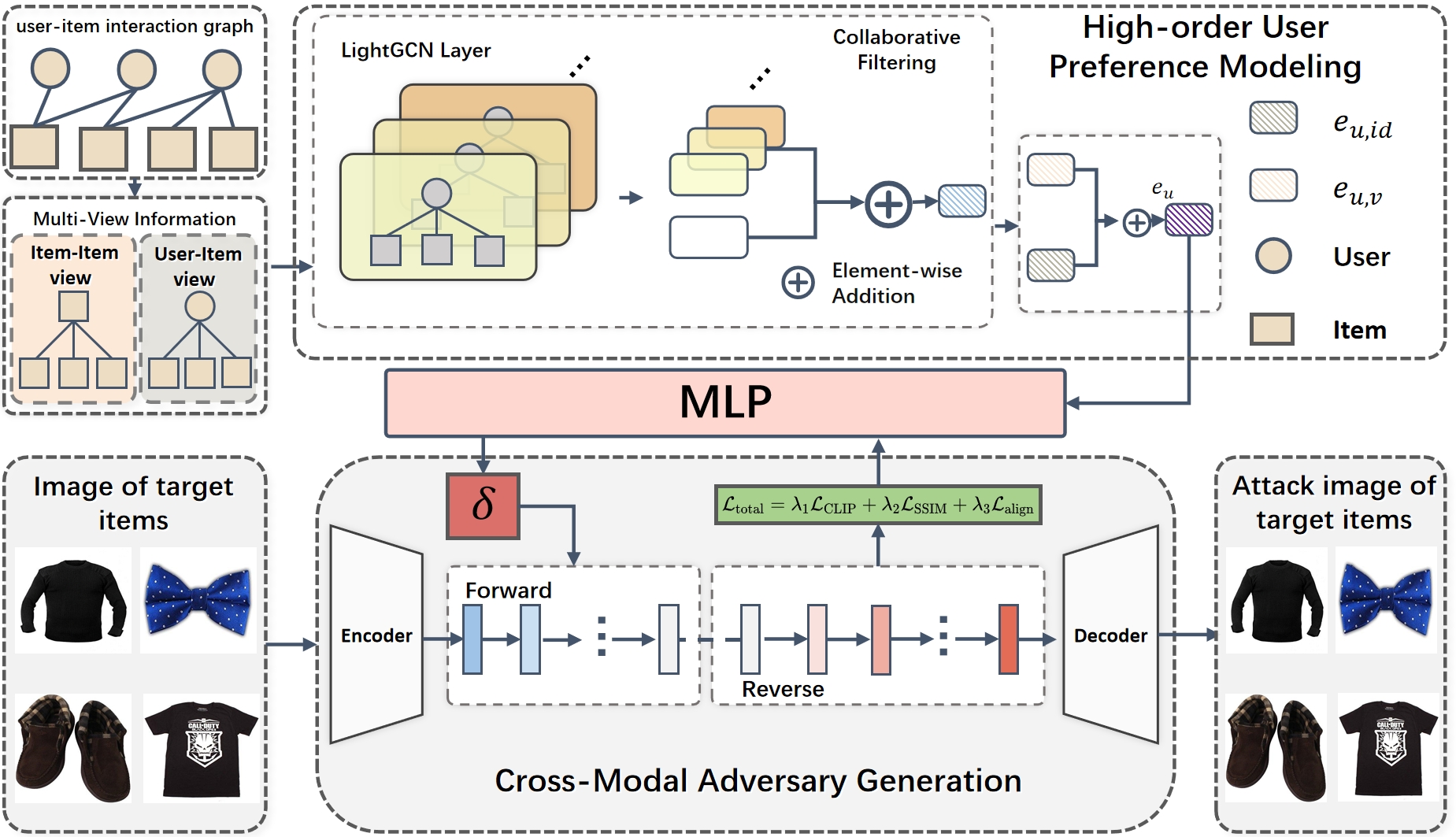}
  \caption{\textbf{AUV-Fusion: A Cross-modal adversarial attack framework for Visual-Aware Recommender Systems.} Our method integrates two key modules: the upper layer leverages a graph convolutional network to model high-order user preferences by capturing collaborative interactions and visual semantics, while the lower layer utilizes a diffusion-based adversary generation process that injects user-driven perturbations into the latent space.}
  \label{fig:framework} 
\end{figure*}
We present AUV-Fusion, a novel adversarial attack framework 
that synergizes graph-based high-order user preference modeling with 
diffusion-driven perturbations. As illustrated in Figure~\ref{fig:framework}, the architecture consists of two cohesive components: 1) the upper high-order user preference modeling, and 
2) the lower Cross-modal Adversary Generation connected through MLP-based feature projection.

\subsection{High-order User Preference Modeling}\label{sec:HoU}

We leverage a Graph Convolutional Network (GCN) to extract deep features from the user–item interaction graph \(G\), capturing both collaborative signals and semantic visual features that are essential for multimedia recommendation~\cite{yu2023multi,wei2024llmrec,xu2024mentor}. To this end, we construct two complementary views. In the user–item view, under the LightGCN~\cite{he2020lightgcn} framework, multi-layer graph convolutions are performed on \(G\). User and item embeddings are updated by aggregating features from their neighbors at each layer, where the initial embeddings \(\mathbf{e}_{u,id}^{(0)}\) and \(\mathbf{e}_{i,id}^{(0)}\) are randomly initialized and trainable. The propagation at the \(k\)-th layer is given by
\begin{equation}
\left\{\begin{aligned}
    \mathbf{e}_{u,\text{id}}^{(k)} &= \sum_{i \in N_u} \frac{1}{\sqrt{|N_u|} \sqrt{|N_i|}} \mathbf{e}_{i,\text{id}}^{(k-1)}, \\
\mathbf{e}_{i,\text{id}}^{(k)} &= \sum_{u \in N_i} \frac{1}{\sqrt{|N_i|} \sqrt{|N_u|}} \mathbf{e}_{u,\text{id}}^{(k-1)},
\end{aligned}
\right.
\end{equation}
with \(N_u\) and \(N_i\) representing the neighboring items and users, respectively; symmetric normalization mitigates node degree imbalance. By aggregating the layer-wise information, the final user–item representation is defined as
\begin{equation}
\mathbf{e}_{u,\text{id}} = \frac{1}{K+1}\sum_{k=0}^{K} \mathbf{e}_{u,\text{id}}^{(k)}.
\end{equation}

In parallel, the item–item view enriches item features through an affinity graph constructed from raw image features. We first compute the cosine similarity between item feature vectors:
\begin{equation}
s_{v}^{a,b} = \frac{(\mathbf{e}_{v}^{a})^T \mathbf{e}_{v}^{b}}{\|\mathbf{e}_{v}^{a}\| \|\mathbf{e}_{v}^{b}\|},
\end{equation}
and then sparsify this fully-connected graph using a KNN approach~\cite{yu2023multi}, where for each item \(a\) only the top-\(K\) most similar edges are retained:
\begin{equation}
\tilde{s}_{v}^{a,b} =
\begin{cases}
s_{v}^{a,b}, & \text{if } s_{v}^{a,b} \in \operatorname{top}\text{-}K(\{ s_{a,c} : c \in I\}), \\
0, & \text{otherwise}.
\end{cases}
\end{equation}
After sparsification, we normalize the affinity matrix to mitigate the impact of high-degree nodes. Specifically, letting \(D_v\) be the diagonal degree matrix of \(\tilde{S}_v\), the normalized affinity matrix is computed as
\begin{equation}
\hat{\mathbf{S}}_v = \mathbf{D}_v^{-1/2} \tilde{\mathbf{S}}_v \mathbf{D}_v^{-1/2}.
\end{equation}
This normalization balances the propagation of item visual features. We then perform graph convolution on the raw item visual features \(\mathbf{E}_{i,v}\) using \(\hat{\mathbf{S}}_v\), which refines and enhances the semantic representations across connected items. Finally, for each user \(u\), we aggregate the enhanced visual features from the items they interacted with by applying a symmetric normalization, resulting in the final visual feature:
\begin{equation}
\mathbf{e}_{u,\text{v}} = \sum_{i \in N_u} \frac{1}{\sqrt{|N_u| \cdot |N_i|}} \mathbf{e}_{i,\text{v}}.
\end{equation}
The enhanced behavioral and visual features are fused to form the comprehensive user preference representation:
\begin{equation}
\mathbf{e}_u = \mathbf{e}_{u,\text{id}} + \mathbf{e}_{u,\text{v}}.
\end{equation}

\subsection{Cross-modal Adversary Generation}

In this section, we present our cross-modal adversary generation framework, which integrates authentic user interaction data to guide the production of semantically consistent adversarial perturbations. The framework consists of two major components: the transformation of user embeddings into a latent-space perturbation, and the incorporation of this perturbation into a diffusion-based image generation process. The following subsections detail these components and the associated optimization strategy.

\subsubsection{Latent Space Perturbation Generation}\label{sec:latent-perturbation}

To generate a perturbation that reflects a target user's preferences, we first obtain the aggregated user embedding \( \mathbf{e}_u \) (as described in Section~\ref{sec:HoU}). This embedding is mapped to a latent-space perturbation via a multilayer perceptron (MLP) \(\mathcal{P}\) as follows:
\( \mathbf{\delta} = \mathcal{P}(\mathbf{e}_u), \) where \(\delta \in \mathbb{R}^{4 \times 28 \times 28}\) is dimensioned to match the latent space of the Variational Autoencoder (VAE) used in our diffusion model. Our MLP \(\mathcal{P}\) operates in three integrated steps. First, the raw user embedding \( \mathbf{e}_u \in \mathbb{R}^{d} \) is enhanced by a fully-connected layer that projects it into a 256-dimensional space, yielding
\begin{equation}
\mathbf{x} = \mathrm{BN}(\mathrm{LeakyReLU}(\mathbf{W}_1 \mathbf{e}_u + \mathbf{b}_1)),
\end{equation}
where Batch Normalization and a LeakyReLU activation (with a negative slope of 0.2) serve to deepen the feature representation. Next, these enhanced features are aggregated using multi-head attention. Specifically, we compute a global query \( \mathbf{q} = \mathrm{mean}(\mathbf{x}) \) and employ it to attend over \(\mathbf{x}\), resulting in the aggregated representation
\begin{equation}
\mathbf{z} = \mathrm{Attention}(\mathbf{q}, \mathbf{x}, \mathbf{x}) \in \mathbb{R}^{256},
\end{equation}
Finally, \( \mathbf{z} \)is transformed through a two-layer fully-connected network: first, we apply a ReLU activation to obtain \( \mathbf{y} = \mathrm{ReLU}(\mathbf{W}_2 \mathbf{z} + \mathbf{b}_2) \), then use a Tanh function to produce the vector \( \delta_{\mathrm{vec}} = \mathrm{Tanh}(\mathbf{W}_3 \mathbf{y} + \mathbf{b}_3) \), which is reshaped into the perturbation tensor
\begin{equation}
\mathbf{\delta} = \mathrm{reshape}(\mathbf{\delta}_{\text{vec}}, [4, 28, 28]).
\end{equation}
This integrated design ensures that the generated perturbation \(\delta\) is well-aligned with the latent representation of our diffusion model, enabling effective adversarial modifications driven by actual user preferences.

\subsubsection{Diffusion Process with Adversarial Injection}

After obtaining the latent-space perturbation \( \mathbf{\delta} \), we integrate it into the diffusion process to generate an adversarial image that reflects user-specific modifications. First, the input image \( x_0 \) is preprocessed and encoded by a pretrained VAE to yield its latent representation \( \mathbf{z}_0 = \mathrm{VAE_{enc}}(x_0) \), compressing the image into a lower-dimensional space that preserves essential semantic features. Next, Gaussian noise is injected into \( \mathbf{z}_0 \) following a diffusion schedule, such that at time step \( t \) the noisy latent is computed as 
\begin{equation}
\mathbf{z}_t = \sqrt{\bar{\alpha}_t}\, \mathbf{z}_0 + \sqrt{1 - \bar{\alpha}_t}\, \epsilon, \quad \epsilon \sim \mathcal{N}(0, I),
\end{equation}
with \( \bar{\alpha}_t = \prod_{i=1}^t \alpha_i \) controlling the cumulative noise. At an appropriate step, we scale the adversarial perturbation by a factor \( \eta \) and inject it into the noisy latent, resulting in \( \tilde{\mathbf{z}}_t = \mathbf{z}_t + \eta \cdot \mathbf{\delta}, \) Thereafter, the reverse diffusion process employs deterministic DDIM~\cite{Rombach_2022_CVPR} sampling to reconstruct the adversarial image. At each time step \( t \), the UNet predicts the noise component via: \( \hat{\epsilon}_t = \mathrm{UNet}(\tilde{\mathbf{z}}_t, t, \mathbf{h}),\) where the intentionally empty text embedding \( \mathbf{h} \) is used so that no additional textual guidance interferes with the influence of \( \mathbf{e}_u \). The latent is iteratively updated:
\begin{equation}
\tilde{\mathbf{z}}_{t-1} = \sqrt{\alpha_{t-1}}\, \left(\frac{\tilde{\mathbf{z}}_t - \sqrt{1-\alpha_t}\,\hat{\epsilon}_t}{\sqrt{\alpha_t}}\right) + \sqrt{1-\alpha_{t-1}}\, \hat{\epsilon}_t,
\end{equation}
and finally, the adversarial image is reconstructed by decoding the final latent: \(x_{\mathrm{adv}} = \mathrm{VAE_{dec}}(\tilde{\mathbf{z}}_0).\)

To ensure the generated adversarial image retains high visual fidelity while incorporating effective user-specific modifications, we define a composite loss function comprising three terms. The CLIP loss,
\begin{equation}
\mathcal{L}_{\mathrm{CLIP}} = \|\mathrm{CLIP}(x_0) - \mathrm{CLIP}(x_{\mathrm{adv}})\|_2^2,
\end{equation}
preserves high-level semantic features; the SSIM loss,
\begin{equation}
\mathcal{L}_{\mathrm{SSIM}} = 1 - \mathrm{SSIM}(x_0, x_{\mathrm{adv}}),
\end{equation}
ensures structural similarity; and the user alignment loss,
\begin{equation}
\mathcal{L}_{\mathrm{align}} = -\,\mathrm{cosine}(f(x_{\mathrm{adv}}), \mathbf{e}_u),
\end{equation}
aligns the adversarial image with genuine user preferences by projecting image features into the user embedding space. The feature extractor \(f\) is instantiated with an off‑the‑shelf pretrained model from the torchvision library (e.g.\ ResNet) and employed without any further fine‑tuning, in accordance with standard black‑box attack methodology to minimize computational overhead. The overall loss is defined as
\begin{equation}
\mathcal{L}_{\mathrm{total}} = \lambda_1 \mathcal{L}_{\mathrm{CLIP}} + \lambda_2 \mathcal{L}_{\mathrm{SSIM}} + \lambda_3 \mathcal{L}_{\mathrm{align}}.
\end{equation}
which is minimized via gradient descent to train the MLP module. In this manner, our framework systematically embeds a user-driven adversarial perturbation into the diffusion process, ultimately generating an adversarial image that is visually coherent, semantically faithful, and tailored to the target user's preferences.
\section{Experiments}
In this section, we present comprehensive experiments designed to evaluate the performance of our proposed method against several baselines. Our experimental study is organized around the following four research questions:

\begin{itemize}[leftmargin=*]
    \item \textbf{RQ1: Performance Comparison.}  
    How does our method compare with state-of-the-art approaches in terms of effectiveness, particularly with respect to \emph{Exposure Gain}?

    \item \textbf{RQ2: Concealment Evaluation.}  
    How well does our method achieve the concealment of monitored risks compared to alternative methods, as measured by \emph{Detection Risk}?

    \item \textbf{RQ3: Impact of User Preferences.}  
    What role do user interaction data and preferences play in enhancing the effectiveness of our method? 

    \item \textbf{RQ4: Hyperparameter Sensitivity.}  
    How do different hyperparameters affect the performance and robustness of our method? 
\end{itemize}

\subsection{Experiment Setup}
\begin{wraptable}{r}{0.4\linewidth}
  \centering
    \vspace{-7.5mm}
    \caption{The statistics of datasets.}
    \vspace{2mm}
  \begin{adjustbox}{width=\linewidth}
  \label{tab:1}
  \begin{tabular}{c|ccc}
    \toprule
    Dataset & \#Users & \#Items & \#Interactions \\
    \midrule
    Amazon Men & 34,244 & 110,636 & 254,870 \\
    Tradesy.com & 33,864 & 326,393 & 655,409 \\
    \bottomrule
  \end{tabular}
  \end{adjustbox}
    \vspace{-4mm}
\end{wraptable}

\textbf{Datasets selection.}
To strictly evaluate the effectiveness of AUV-Fusion, we selected two real datasets with relatively complete data and publicly accessible, namely the Amazon Men's Clothing dataset and the Tradesy.com dataset~\cite{he2016vbpr}~\cite{kang2017visually}~\cite{liu2021adversarial}. As shown in Table~\ref{tab:1}, there are significant differences between the two in terms of data volume and interaction quantity. This difference ensures that our evaluation covers a wide range of fashion items and user preferences. The core input of AUV-Fusion is composed of descriptive images of each product. Following the related work in ~\cite{kang2017visually}~\cite{liu2021adversarial},we convert the numerical ratings in the dataset into binary form, which effectively represents them as implicit feedback. Finally, we employed the leave-one-out method~\cite{he2017neural} to partition the training and test sets of both datasets in a 9:1 ratio, ensuring sufficient training data while providing a reasonably sized test set to maintain the robustness and comparability of the experiment.

\textbf{Baseline attack methods.}
We conduct a comprehensive comparison between our proposed \textbf{AUV-Fusion} method and state-of-the-art item representation attack approaches, including (\textbf{INSA}, \textbf{EXPA})~\cite{liu2021adversarial}, and the stylized attack \textbf{SPAF}~\cite{yang2024attacking}. For fair baseline comparisons, we strictly adhere to the hyperparameter configurations recommended in their original publications. In particular, for both \textbf{INSA} and \textbf{EXPA}, we maintain a consistent perturbation budget of $\epsilon = 32$ (within the pixel value range of $[0,255]$) to ensure comparable attack strength, and standardize the training to 100 epochs to guarantee convergence while preserving experimental efficiency.

For \textbf{SPAF}, we implement the following precise configuration: the main attack module is trained for 100 epochs with a batch size of 256 and an objective balancing weight $\lambda = 1$, and a uniform learning rate of $10^{-4}$ is applied across all submodules. The decoder in \textbf{SPAF} is trained with an epoch count of 300,000 and a batch size of 8 for each dataset, which corresponds to approximately 18 hours of training per dataset on an NVIDIA GeForce RTX 4090 GPU. For the StyleBPR component of \textbf{SPAF}, we set the number of epochs to 300, the batch size to 512, and the embedding dimension $d$ to 100, with all learning rates in the SPAF model fixed at 0.0001. 

\textbf{Targeted recommender systems.}\label{TRS}
Inspired by recent research  ~\cite{DMM20,liu2021adversarial}, we assess SPAF's performance using three distinct visual recommendation systems as victim models: the traditional VBPR~\cite{he2016vbpr}, the cutting-edge DVBPR~\cite{kang2017visually}, and AMR~\cite{tang2019adversarial}, which integrates adversarial robustness mechanisms. We adopt the model implementations and parameter settings provided by a public open-source library\footnote{The open-source library is available at \url{https://github.com/liuzrcc/AIP} for further details on implementation and configuration.}. Additionally, we employ a two-stage evaluation framework frequently used in industry ~\cite{10.1145/1864708.1864770,10.1145/3219819.3219869}. In the first stage, a BPR model is applied to shortlist relevant candidate items, and in the subsequent stage, one of the victim models is used to generate the final ranking.

\textbf{Configuration of AUV-Fusion.}
In the cross-modal adversary generation layer, a diffusion model with 50 steps is integrated. Noise amplitude is adaptively regulated via a standard deviation parameter, while parameters unrelated to the diffusion process remain frozen for stability. Notably, an MLP generates the adversarial perturbations and is trained for 100 epochs during the attack phase, ensuring efficient adversarial adaptation without compromising the semantic integrity of the generated images.

\subsection{Exposure Gain Analysis (RQ1)}
{
\begin{table*}[t]
    \caption{\textbf{HR@k Comparison of AUV-Fusion, EXPA, INSA and SPAF Attacks ($k \in \{5,10,20\}$) under a Black-Box Setting (\textit{i.e.}, $p = 0.1$).} For each visually-aware recommender system and dataset, the best result is highlighted in bold and the second-best is underlined. A higher HR@k indicates superior attack performance by reflecting a greater likelihood of the target item appearing in the top-$k$ recommendations.}
    \vspace{2mm}
    \label{tab:2}
    \resizebox{\textwidth}{!}{
    \footnotesize
    \centering
    \begin{tabular}{c|c|c|c|c|c|c|c}
        \toprule
        \multirow{2}{*}{Victim VARS} & \multirow{2}{*}{Attack Method} & \multicolumn{3}{c|}{Amazon Men} & \multicolumn{3}{c}{Tradesy.com} \\
        \cmidrule{3-8} 
        & & HR@5  & HR@10  & HR@20 & HR@5  & HR@10  & HR@20  \\
        \midrule
        \multirow{4}{*}{VBPR} 
        & EXPA & 0.00021$_{\pm 0.00001}$ & 0.00064$_{\pm 0.00004}$ & 0.00201$_{\pm 0.00007}$ & 0.00193$_{\pm 0.00009}$ & 0.00439$_{\pm 0.00014}$ & 0.00986$_{\pm 0.00032}$ \\
        & INSA & 0.00117$_{\pm 0.00009}$ & 0.00264$_{\pm 0.00014}$ & 0.00660$_{\pm 0.00029}$ & 0.00265$_{\pm 0.00004}$ & 0.00574$_{\pm 0.00005}$ & 0.01218$_{\pm 0.00007}$ \\
        & SPAF & \underline{0.00241}$_{\pm 0.00088}$ & \underline{0.00526}$_{\pm 0.00175}$ & \underline{0.01193}$_{\pm 0.00327}$ & \textbf{0.00354}$_{\pm 0.00008}$ & \textbf{0.00710}$_{\pm 0.00010}$ & \textbf{0.01371}$_{\pm 0.00016}$ \\
        \cmidrule{2-8}
        & \textbf{AUV-Fusion} & \textbf{0.02047}$_{\pm 0.00002}$ & \textbf{0.02995}$_{\pm 0.00002}$ & \textbf{0.04511}$_{\pm 0.00003}$ & \underline{0.00347}$_{\pm 0.00000}$ & \underline{0.00693}$_{\pm 0.00001}$ & \underline{0.01368}$_{\pm 0.00001}$ \\
        \midrule
        \multirow{4}{*}{DVBPR} 
        & EXPA & 0.00024$_{\pm 0.00010}$ & 0.00074$_{\pm 0.00021}$ & 0.00221$_{\pm 0.00040}$ & 0.00139$_{\pm 0.00035}$ & 0.00285$_{\pm 0.00046}$ & 0.00611$_{\pm 0.00057}$ \\
        & INSA & 0.00028$_{\pm 0.00004}$ & 0.00089$_{\pm 0.00012}$ & 0.00257$_{\pm 0.00030}$ & 0.00147$_{\pm 0.00014}$ & 0.00314$_{\pm 0.00023}$ & 0.00658$_{\pm 0.00036}$ \\
        & SPAF & \underline{0.00061}$_{\pm 0.00011}$ & \underline{0.00186}$_{\pm 0.00021}$ & \underline{0.00562}$_{\pm 0.00049}$ & \underline{0.00175}$_{\pm 0.00032}$ & \underline{0.00359}$_{\pm 0.00049}$ & \underline{0.00729}$_{\pm 0.00083}$ \\
        \cmidrule{2-8}
        & \textbf{AUV-Fusion} & \textbf{0.00109}$_{\pm 0.00017}$ & \textbf{0.00276}$_{\pm 0.00023}$ & \textbf{0.00632}$_{\pm 0.00035}$ & \textbf{0.00204}$_{\pm 0.00044}$ & \textbf{0.00386}$_{\pm 0.00054}$ & \textbf{0.00756}$_{\pm 0.00074}$ \\
        \midrule
        \multirow{4}{*}{AMR} 
        & EXPA & 0.00179$_{\pm 0.00003}$ & 0.00241$_{\pm 0.00003}$ & 0.00332$_{\pm 0.00005}$ & 0.00091$_{\pm 0.00001}$ & 0.00233$_{\pm 0.00001}$ & 0.00577$_{\pm 0.00005}$ \\
        & INSA & 0.00372$_{\pm 0.00005}$ & 0.00494$_{\pm 0.00006}$ & 0.00667$_{\pm 0.00006}$ & 0.00124$_{\pm 0.00001}$ & 0.00291$_{\pm 0.00003}$ & 0.00690$_{\pm 0.00003}$ \\
        & SPAF & \underline{0.00488}$_{\pm 0.00053}$ & \underline{0.00638}$_{\pm 0.00067}$ & \underline{0.00865}$_{\pm 0.00091}$ & \underline{0.00145}$_{\pm 0.00003}$ & \underline{0.00344}$_{\pm 0.00008}$ & \underline{0.00760}$_{\pm 0.00013}$ \\
        \cmidrule{2-8}
        & \textbf{AUV-Fusion} & \textbf{0.00946}$_{\pm 0.00001}$ & \textbf{0.01175}$_{\pm 0.00000}$ & \textbf{0.01518}$_{\pm 0.00000}$ & \textbf{0.00201}$_{\pm 0.00000}$ & \textbf{0.00440}$_{\pm 0.00000}$ & \textbf{0.00977}$_{\pm 0.00001}$ \\
        \bottomrule
    \end{tabular}
    }
\end{table*}
}    
{\footnotesize
\begin{table}[t]
    \centering
    \caption{\textbf{HR@k results of cross-system attack performance on VBPR (\textit{i.e.}, $p = 0$).} Attacks are trained on one real-world dataset and tested on VBPR models trained on a different dataset.}
    \label{tab:3}
    \resizebox{0.8\textwidth}{!}{

    \begin{tabular}{c|c|c|c|c}
        \toprule
        Attack $\to$ VARS & Attack & H@5 & H@10 & H@20 \\
        \hline
        \multirow{4}{*}{Tradesy $\to$ Amazon} 
        & EXPA & 0.00020$_{\pm 0.00002}$ & 0.00065$_{\pm 0.00003}$ & 0.00204$_{\pm 0.00009}$ \\
        & INSA & 0.00130$_{\pm 0.00008}$ & 0.00285$_{\pm 0.00013}$ & 0.00707$_{\pm 0.00030}$ \\
        & SPAF & \underline{0.00294}$_{\pm 0.00071}$ & \underline{0.00645}$_{\pm 0.00149}$ & \underline{0.01476}$_{\pm 0.00284}$ \\
        \cmidrule{2-5}
        & AUV-Fusion & \textbf{0.01881}$_{\pm 0.00040}$ & \textbf{0.02822}$_{\pm 0.00045}$ & \textbf{0.04327}$_{\pm 0.00052}$ \\
        \hline
        \multirow{4}{*}{Amazon $\to$ Tradesy} 
        & EXPA & 0.00196$_{\pm 0.00008}$ & 0.00448$_{\pm 0.00015}$ & 0.00994$_{\pm 0.00026}$ \\
        & INSA & 0.00277$_{\pm 0.00005}$ & 0.00587$_{\pm 0.00009}$ & 0.01247$_{\pm 0.00014}$ \\
        & SPAF & \textbf{0.00364}$_{\pm 0.00003}$ & \textbf{0.00721}$_{\pm 0.00007}$ & \textbf{0.01400}$_{\pm 0.00006}$  \\
        \cmidrule{2-5}
        & AUV-Fusion & \underline{0.00354}$_{\pm 0.00004}$ & \underline{0.00701}$_{\pm 0.00005}$ & \underline{0.01345}$_{\pm 0.00010}$ \\
        \bottomrule
    \end{tabular}
    }
\end{table}
}

Table~\ref{tab:2} and Table~\ref{tab:3} present our evaluation results, focusing on the overall exposure gain achieved by the proposed attack methods.

\textbf{Standard exposure gain analysis.}
Table~\ref{tab:2} shows the HR@k performance for various attack methods evaluated under a black-box setting on different visually-aware recommender systems and datasets (Amazon Men and Tradesy.com). Our method, AUV-Fusion, consistently outperforms the baselines, with marked improvements especially evident in scenarios involving more complex architectures such as DVBPR and AMR. 

The results demonstrate that by incorporating genuine user preferences into the generation of adversarial perturbations, AUV-Fusion achieves a significant boost in exposure rates. In the Amazon Men setting for VBPR, for example, the HR@5 obtained by AUV-Fusion is orders of magnitude higher than those of both EXPA and INSA, clearly reflecting its superior capability in highlighting target items. Although SPAF exhibits a slight edge on certain metrics for VBPR on the Tradesy.com dataset, the overall performance of AUV-Fusion remains robust across different models. AMR, fortified with robust adversarial training defenses, clearly demonstrates the superiority of our user-guided approach. Despite its advanced defense mechanisms, AUV-Fusion delivers notable improvements, indicating that the integration of high-order user preference data enables the generation of adversarial images that remain visually coherent and closely aligned with genuine user behavior.

Collectively, these findings affirm that our cross-modal attack framework, which systematically utilizes user preference data, offers a substantial improvement over existing methods that rely solely on visual cues.

\textbf{Cross-domain exposure gain analysis.}
Table~\ref{tab:3} presents the HR@k metrics for cross-domain attack performance on the VBPR model, where attacks are trained on one real-world dataset and evaluated on VBPR models trained on a different dataset. When transferring attacks from Tradesy to Amazon, AUV-Fusion achieves significantly higher HR@k values (0.01881 at HR@5, 0.02822 at HR@10, and 0.04327 at HR@20) compared to the baselines. This result demonstrates that leveraging genuine user interaction signals allows AUV-Fusion to generate effective adversarial perturbations even when the training and testing domains differ markedly.
In contrast, in the Amazon-to-Tradesy scenario, SPAF attains the best performance, with AUV-Fusion ranking second. This observation suggests that the visual characteristics present in the Amazon-trained model may be more conducive to the style-transfer mechanism underpinning SPAF, whereas the strength of our method lies in its ability to transfer adversarial strength effectively when substantial domain differences exist. 

\subsection{Attack Imperceptibility Analysis (RQ2)}
To comprehensively evaluate the detection risk $\mathcal{D}(\cdot)$ of our adversarial attacks, we assess their imperceptibility from two complementary perspectives: (i) Overall imperceptibility via human user studies, and (ii) Semantic imperceptibility through feature-level analysis.

\begin{figure}
    \begin{minipage}{0.445\linewidth}
        \centering
      \includegraphics[width=0.9\linewidth]{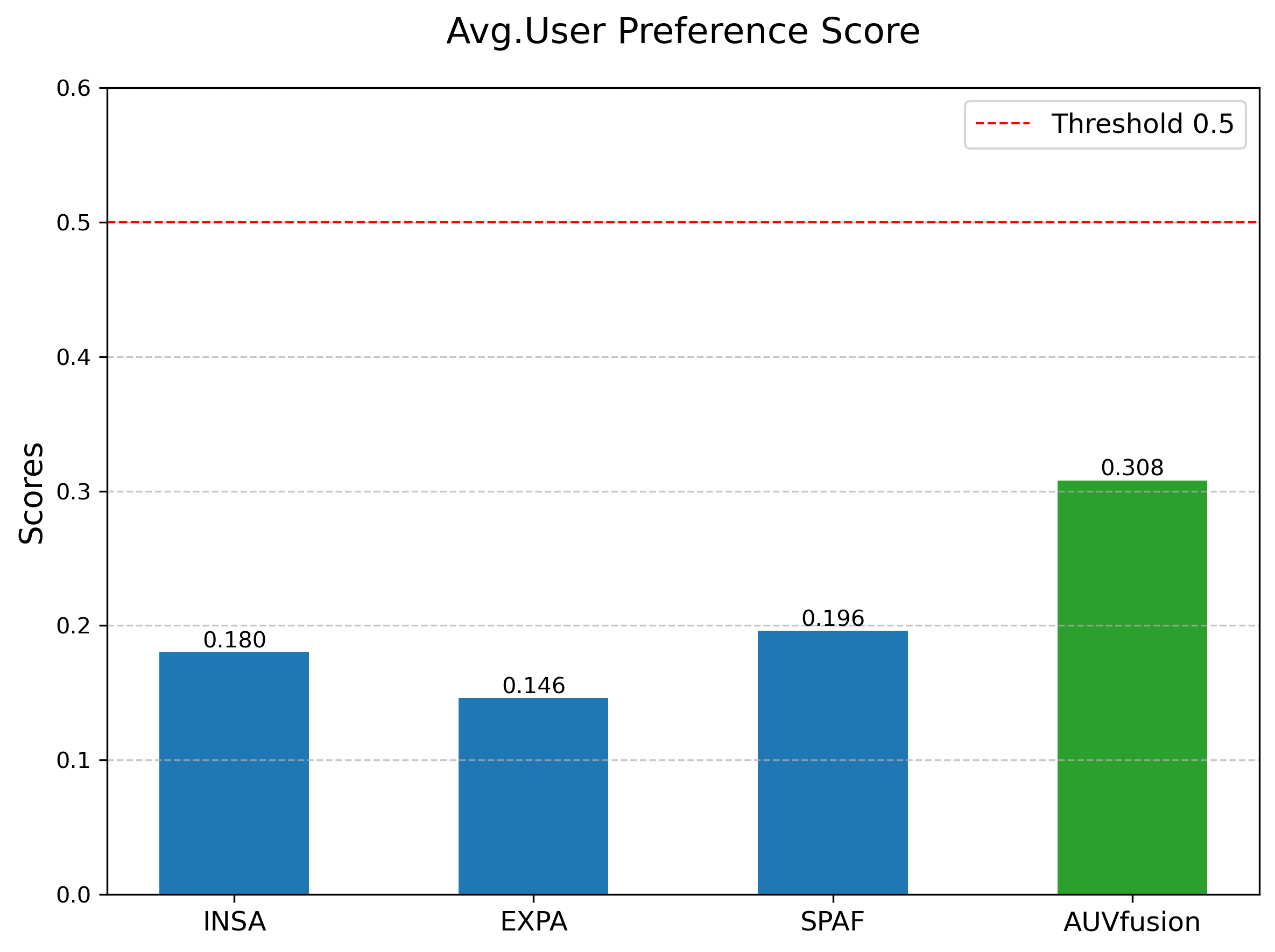}
      \caption{A human study shows that a value close to 0.5 indicates that users cannot distinguish between adversarial images and benign images (the adversarial samples are very realistic).}
      \label{human}
    \end{minipage}
    \hfill
    \begin{minipage}{0.543\linewidth}
        \centering
      \includegraphics[width=\linewidth]{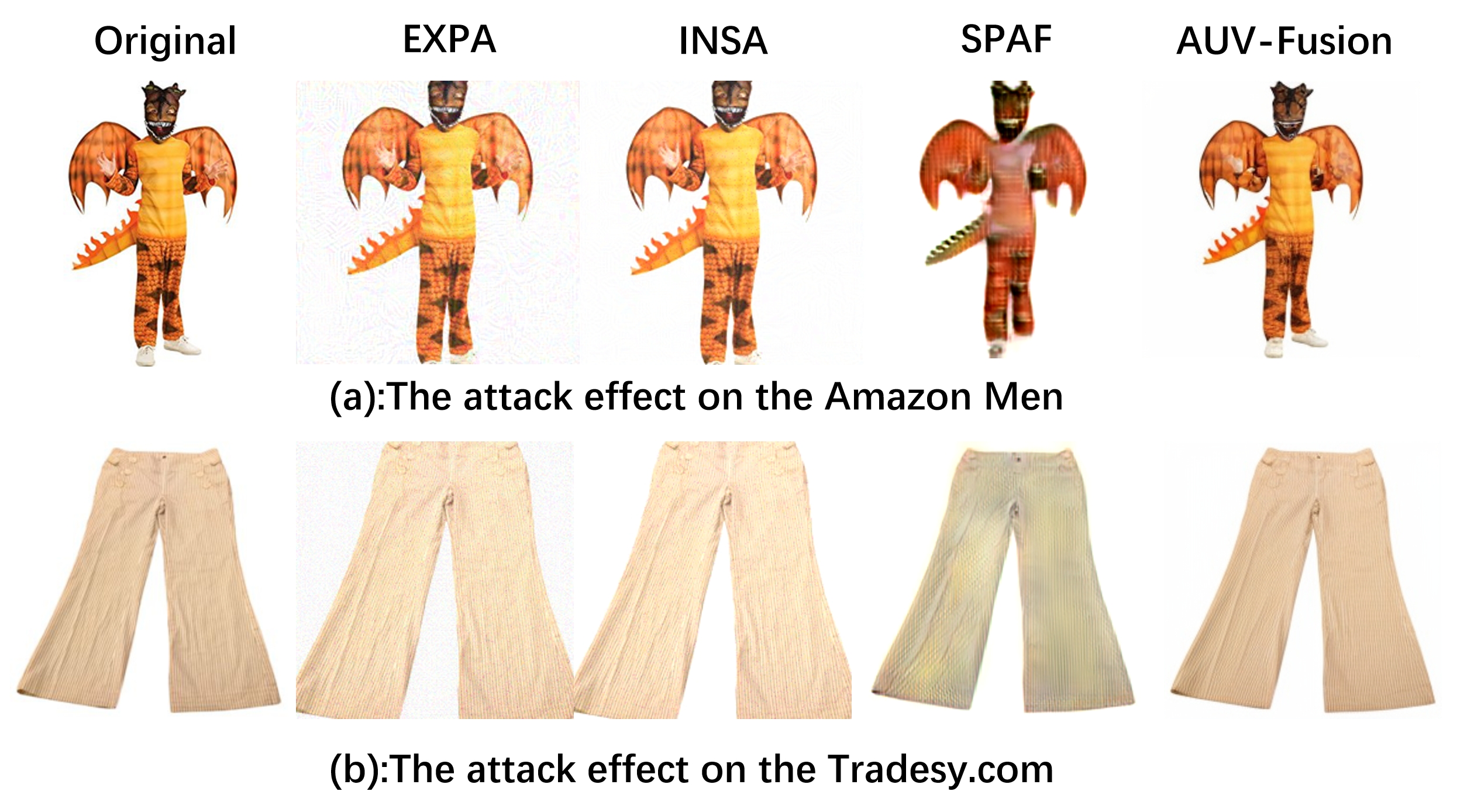}
      \caption{\textbf{The color contrast of the images and the effectiveness of different attack methods on two datasets.} The first column shows the original images, and the far right column shows the adversarial images after the AUV-Fusion attack.}
      \label{fig:comparison2} 
    \end{minipage}
\end{figure}

\textbf{Overall imperceptibility.}
We conducted a human study inspired by ~\cite{Bhattad2020Unrestricted, yang2024attacking, 10.1007/978-3-319-46487-9_40} to assess the visual quality of the adversarial images generated by AUV-Fusion. In this study, 100 voluntary participants were presented with 50 pairs of images—each pair containing one original image and one generated adversarial image—with 25 pairs sourced from Amazon and 25 pairs from Tradesy. Participants were instructed to identify the original image in each pair. Each pair was displayed for 5 seconds to reflect the typical product viewing duration on e-commerce platforms, and participants were allowed 5 practice trials prior to the main evaluation. The user preference score for each image was calculated as the ratio of the number of times it was selected to the number of times it was displayed; with a score of 0.5 suggesting complete indistinguishability between original and adversarial samples. As shown in Figure~\ref{human} and~\ref{fig:comparison2}, AUV-Fusion has a higher imperceptibility than other attack methods.

\begin{figure}
    \begin{minipage}{0.538\linewidth}
        \centering
          \includegraphics[width=\linewidth]{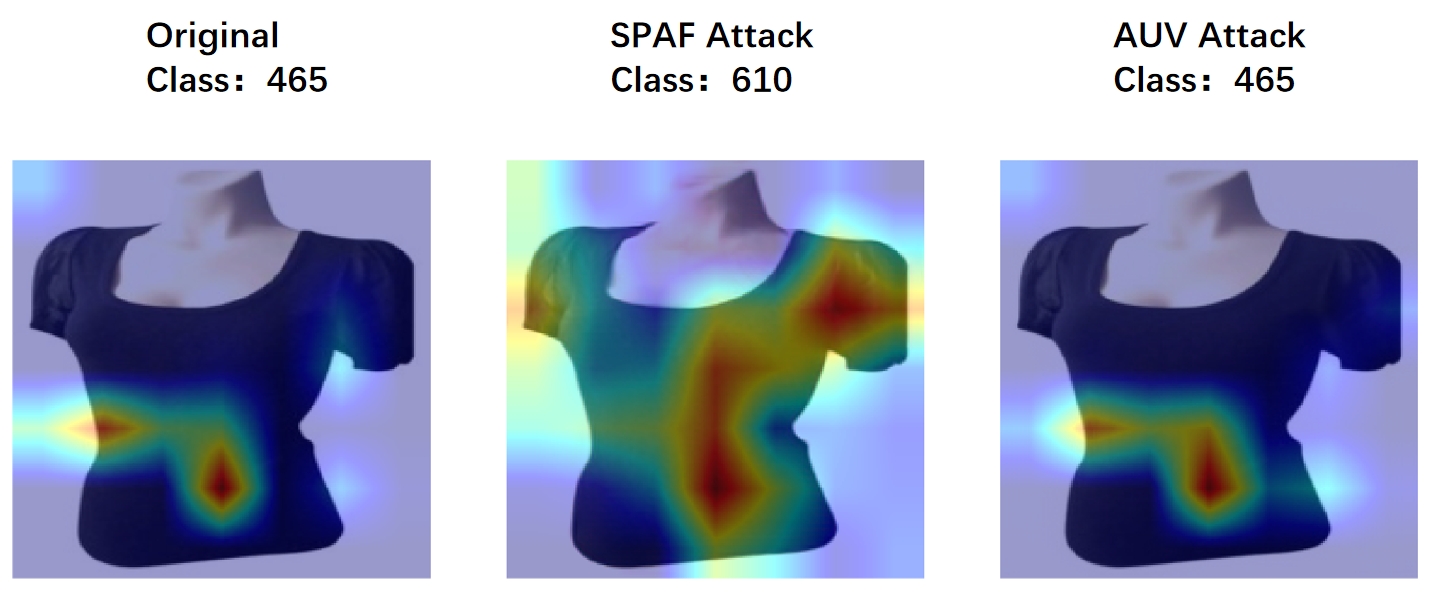}
          \caption{Semantic heatmaps comparing adversarial images after SPAF attacks and AUV-Fusion attacks with the original images.}
          \label{p1}
    \end{minipage}
    \hfill
    \begin{minipage}{0.45\linewidth}
        \centering
  \includegraphics[width=\linewidth]{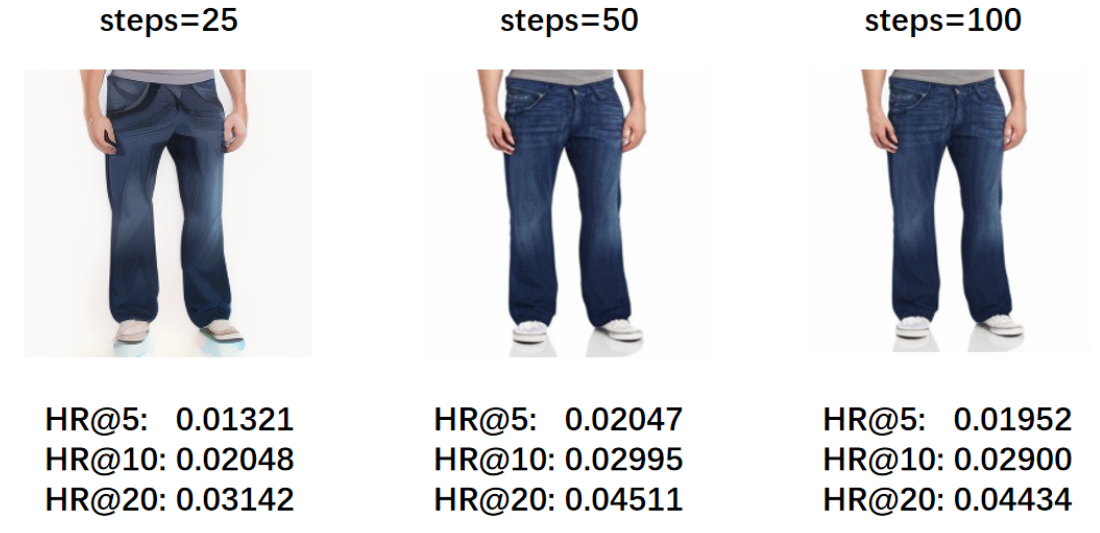}
  \caption{The influence of diffusion steps on the VBPR of the AUV-Fusion model on the Amazon.}
  \label{steps}
    \end{minipage}
\end{figure}

\textbf{Semantic imperceptibility.}
Evaluating semantic consistency is crucial for assessing the degree of concealment of adversarial modifications\cite{10.1145/3674399.3674448}. Figure~\ref{p1} presents a comparison of semantic heatmaps \cite{DBLP:journals/corr/SelvarajuDVCPB16,10.5555/3454287.3454299}: the left image shows the original image, the middle image shows the adversarial sample generated by the SPAF attack, and the right image shows the semantic heatmap derived from the AUV-Fusion attack. The SPAF attack modifies the semantic content through transformations in the target feature space, resulting in obvious semantic changes. However, the AUV-Fusion attack restricts its perturbations within the local neighborhood of the original data distribution, thereby reducing the semantic modification to the original image. Although both methods maintain visual naturalness, AUV-Fusion attack preserves semantic consistency, thereby reducing the risk of being detected at the feature level.

\subsection{Ablation Study (RQ3)}
\begin{table}[t]
    \centering
    \setlength{\tabcolsep}{3pt}
    \caption{\textbf{Ablation study on key components of AUV·Fusion on VBPR.}  Variant A (w/o A) omits the user alignment loss, and Variant B (w/o B) disables adversarial injection by setting \( \eta \) to zero.}
    \label{tab:4}
    \resizebox{0.6\textwidth}{!}{
    \centering
    \begin{tabular}{c|c|c|c}
        \toprule
        \multirow{2}{*}{Variants} & \multicolumn{3}{c}{Amazon Men} \\
        \cmidrule{2-4}
        & HR@5 & HR@10 & HR@20\\
        \cmidrule{1-4}
        $\textbf{w/o A}$
        & 0.01084$_{\pm 0.00291}$ & 0.01775$_{\pm 0.00376}$ & 0.02922$_{\pm 0.00456}$  \\
        $\textbf{w/o B}$
        & 0.00818$_{\pm 0.00323}$ & 0.01361$_{\pm 0.00461}$ & 0.02276$_{\pm 0.00609}$   \\
        \cmidrule{1-4}
        $\textbf{AUV-Fusion}$
         & \textbf{0.02047}$_{\pm 0.00002}$ & \textbf{0.02995}$_{\pm 0.00002}$ & \textbf{0.04511}$_{\pm 0.00003}$ \\
        \bottomrule
    \end{tabular}
    }
\end{table}

Table~\ref{tab:4} presents our ablation results on the Amazon Men dataset with the VBPR model. When the user alignment loss is removed (Variant A), the HR@k values drop (\textit{e.g.}, HR@5 decreases from 0.02047 to 0.01084) and the standard deviation increases notably. This suggests that without the loss guiding the adversarial modifications, the attack remains partially effective but becomes much less consistent. Furthermore, disabling the adversarial injection mechanism by setting \( \eta \) to zero (Variant B) results in an even larger decline in HR@k values, accompanied by further increases in variability. These observations indicate that the adversarial injection not only enhances the attack's effectiveness but also contributes to its stability. Overall, these ablation experiments confirm the critical importance of both the user alignment loss and the adversarial injection component in ensuring that AUV-Fusion produces robust, reliable, and user-tailored adversarial images.

\subsection{Hyper-parameter Study (RQ4)}
We investigate the impact of the number of diffusion steps on the adversarial performance of our method using the VBPR model on the Amazon dataset. As shown in Figure~\ref{steps}, when the diffusion steps are set to 25, the HR$@20$ achieves a mean value of approximately 0.03142. Increasing the diffusion steps to 100 results in an HR$@20$ of about 0.04434. Notably, when using 50 diffusion steps, the HR$@20$ reaches 0.04511 with an exceptionally low standard deviation (\(\pm 0.00003\)).

These results suggest that 25 diffusion steps are insufficient for fully integrating the adversarial signal, while 100 steps provide a performance comparable to 50 steps but with potentially higher computational cost or increased variability. The 50-step setting, on the other hand, strikes an optimal balance by maximizing the adversarial exposure gain while ensuring stability in reconstruction. This justifies our choice of 50 diffusion steps in AUV-Fusion.
\section{Conclusion}
In this paper, we proposed AUV-Fusion, a novel adversarial attack framework for Visual-Aware Recommender Systems (VARS), aiming to bridge the gap between adversarial feature manipulation and interaction-driven attacks (\textit{i.e.}, shilling attacks). Our approach leverages a dual-module design, combining a high-order user preference modeling module to capture rich collaborative signals, followed by a cross-modal adversary generation module that uses a diffusion-based process to inject semantically consistent perturbations guided by genuine user interactions.
Our extensive experiments on real-world datasets demonstrate that AUV-Fusion achieves superior exposure gains (notably on complex VARS such as DVBPR) while maintaining low detection risks across both standard and cross-domain scenarios. Compared to state-of-the-art methods like INSA and SPAF, our framework excels in environments where traditional attacks are limited by their single-modal focus and reliance on synthetic manipulations. Overall, AUV-Fusion sets a new benchmark for attack efficacy and stealth in VARS. Future work will explore more advanced multimodal fusion strategies and defense-aware modifications to further enhance the robustness and generalizability of adversarial attacks in recommendation systems.

\bibliographystyle{unsrt}  
\bibliography{references}

\end{document}